\documentclass[twocolumn,superscriptaddress,showpacs,preprintnumbers,amsmath,amssymb,aps,pre]{revtex4}
\usepackage{graphicx}
\usepackage{dcolumn}
\usepackage{bm}
\begin{document}
\title{Is time continuous?}
\author{P. A. Varotsos}
\email{pvaro@otenet.gr} \affiliation{Solid State Section and Solid
Earth Physics Institute, Physics Department, University of Athens,
Panepistimiopolis, Zografos 157 84, Athens, Greece}  

\begin{abstract}
Conventional time is modelled as the one dimensional continuum
$R^1$ of real numbers. This continuity, however, does {\em not} stem
from {\em any} fundamental principle. On the other hand, natural
time is {\em not} continuous and its values as well as those of
the energy, form {\em countable} sets, i.e., with cardinalities
either finite or equal to $\aleph_0$, where this symbol stands for
the {\em transfinite} number of natural numbers. For infinitely
large number of events, the values of natural time form a {\em
denumerable} set, i.e., its cardinality is exactly $\aleph_0$,
while those of conventional time an {\em uncountable} set. This
has a drastically larger cardinality, which in the light of the
 continuum hypothesis becomes equal to $2^{\aleph_0}$.
\end{abstract}
\maketitle

In reviewing the state of physics today, a consensus seems to
emerge that we are missing something absolutely fundamental, e.g.,
\cite{GRO05,HOO05}. Furthermore, there is a widespread belief
that, it is not space but time that in the end poses the greatest
challenge to science (e.g., p.18 of \cite{YOU05}). Time, according
to Weyl (see p.5 of \cite{WEY05}) for example, is ``the primitive
form of the stream of consciousness. It is a fact, however,
obscure and perplexing to our minds, that \dots one does not say
this {\em is} but this is {\em now}, yet no more'' or according to
G\"{o}del ``that mysterious and seemingly self-contradictory being
which, on the other hand, seems to form the basis of the world's
and our own existence.'' (p.111 of \cite{YOU05}). The challenge
seems to stem from the fact that special relativity and quantum
mechanics, which are the two great (and successful) theories of
twentieth-century physics, are based on entirely different ideas,
which are not easy to reconcile (In general, the former theory,
according to Einstein\cite{EIN05}, is an example of ``principled
theory'' in the sense that you start with the principles that
underlie the theory and then work down to deduce the facts,
 while the latter is a ``constructive theory'' meaning that it describes phenomena based
on some known facts but an underlying principle to explain the
strangeness of the quantum world has not yet been found). In
particular, special relativity puts space and time on the same
footing, but quantum mechanics treats them very differently, e.g.,
see p.858 of Ref.\cite{WIL05B}. (In quantum gravity, space is
fluctuating and time is hard to define, e.g., \cite{WIL05A}). More
precisely, as far as the theory of special relativity is concerned, let us
recall the following wording of Einstein\cite{EIN21}:

``Later, H. Minkowski found a particularly elegant and suggestive
expression........, which reveals a formal relationship between
Euclidean geometry of three dimensions and the space time {\em
continuum} of physics........ From this it follows that, in
respect to its {\em role} in the equations of physics, though not
with regard to its physical significance, time is equivalent to
the space co-ordinates (apart from the relations of reality). From
this point of view, physics is, as it were, Euclidean geometry of
four dimensions, or, more correctly, a static in a
four-dimensional Euclidean {\em continuum}.''

whereas in quantum mechanics, Von Neumann complains\cite{VON55}:

``First of all we must admit that this objection points at an
essential weakness which is, in fact, the {\em chief weakness} of
quantum mechanics: its non-relativistic character, which
distinguishes the time t from the three space coordinates x,y,z,
and presupposes an objective simultaneity concept. In fact, while
all other quantities (especially those x,y,z, closely connected
with t by the Lorentz transformation) are represented by
operators, there corresponds to the time an ordinary
number-parameter t, just as in classical mechanics''.

Note also that Pauli\cite{PAU33} has earlier shown that there is
no operator canonically conjugate to the Hamiltonian, if the
latter is bounded from below. This means that for many systems a
time operator does {\em not} exist. In other words, the
introduction of an operator $t$ is basically forbidden and the
time must necessarily be considered as an ordinary number (but
recall the long standing question that Schr\"{o}dinger's equation,
as well as  Einstein's general theory of relativity, is symmetric
under time reversal in contrast to the fact that our world is {\em
not}, e.g., \cite{PEN05}). These observations have led to a quite
extensive literature mainly focused on time-energy (as well as on
``phase-action'') uncertainty relation, proposing a variety of
attempts to overcome these obstacles. The discussion of this
literature, however, lies beyond the scope of the present paper.
We just summarize here that the (conventional) time t is {\em
currently} modelled as the one-dimensional {\em continuum} $R^1$
of the real numbers, e.g., p. 10 of \cite{WIL05A} (or p.12 of
\cite{WEY05} in which it is stated that ``\dots the straight line
\dots is homogeneous and a linear {\em continuum} just like
time''). It is this {\em continuity} on which the present paper is
focused, in a sense that will be explained below.

 It has been recently shown that novel dynamical
features hidden behind time series in complex systems can emerge
upon analyzing them in a new time domain, termed natural
time\cite{VAR01,VAR02,VAR03A,VAR03B,VAR05C,VAR06} (see also below). It seems
that this analysis enables the study of the dynamical {\em evolution} of a complex
system and identifies when the system enters a critical stage.
Hence, natural time may play a key role in predicting impending
catastrophic events in general. Relevant examples of data analysis
in this new time domain have been presented in a large variety of
fields including biology, earth sciences and physics. As a first
example, we mention the analysis of the electrocardiograms which
may herald a cardiac arrest\cite{VAR04,VAR05A}. Secondly, the
detection and the analysis of precursory electric signals, termed
Seismic Electric Signals (e.g.,
\cite{NAT86,VAR86,VAR03C,VARBOOK,APL05}), may
lead\cite{VAR01,VAR02,VAR05C,VAR06} to the prediction of an impending
strong earthquake. A third application of natural time refers to
the manifestation of aging and scaling properties in seismic event
correlation\cite{TIR04, ABE05B}. Finally, as a fourth example we
mention that the data of the avalanches of the
penetration of magnetic flux into thin films of type II
superconductors as well as those of a three dimensional pile of
rice getting progressively closer to the critical state conform to\cite{SAR05}
the features suggested, on the basis of natural time, to describe
critical dynamics.

\begin{figure}
\includegraphics{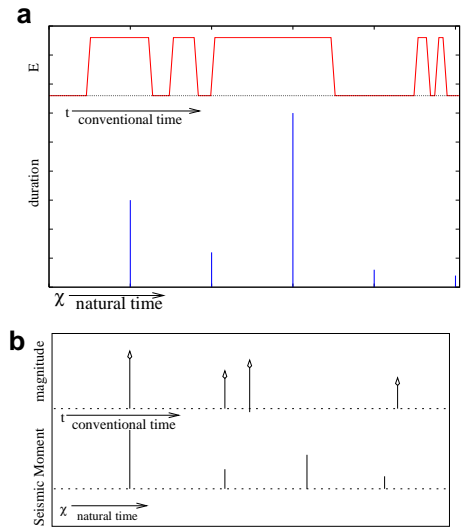}
\caption{\label{fg1} (color online) (a) How a dichotomous series of electric
pulses in conventional time t (upper panel, red) can be read\cite{VAR01,VAR02} in
natural time $\chi$ (lower panel, blue). The symbol $E$ stands for
the electric field ({\em not} to be confused with $E_k$ used in
the text). (b) The same as in (a), but for a series of seismic
events(e.g. \cite{VAR01,VAR05C,VAR06}).}
\end{figure}

In a time series comprising $N$ events, the {\em natural time}
$\chi_k = k/N$ serves as an index
 for the occurrence of the $k$-th event\cite{VAR01,VAR02}, and
 it is smaller than, or equal to, unity (cf. the symbol $\chi$ originates
 from the ancient Greek word $\chi \rho${\' o}$\nu o \varsigma$ (chronos) which means ``time'').
In natural time analysis the evolution of the pair of two
quantities ($\chi_k, E_k$) is considered, where $E_k$ denotes in
general a quantity proportional to the energy of the individual event.
For example, to perform the analysis of seismic events
(Fig.\ref{fg1}(b)), we consider the time evolution of the pair
($\chi_k,M_{0_k})$ where $M_{0_k}$ stands for the seismic moment
of the $k$th event\cite{VAR01, VAR05C, VAR06}, since $M_{0_k}$ is
proportional to the energy emitted in that earthquake (cf.
$M_{0_k}$  differs essentially from the magnitude $M$, but they
are interconnected). As another example, we refer to the analysis
of dichotomous electric signals (Fig.\ref{fg1}(a)) where we
consider $E_k$ as being proportional to the duration of the $k$th
pulse\cite{VAR02,VAR03A,VAR03B}. For the purpose of analysis, the
following {\em continuous} function $\Phi(\omega )$ was
introduced\cite{VAR01,VAR02}:
\begin{equation}
\Phi(\omega)=\frac{\sum_{k=1}^{N} E_{k} \exp \left( i \omega
\frac{k}{N} \right)}{\sum_{n=1}^{N} E_{n}}= \sum_{k=1}^{N} p_k
\exp \left( i \omega \frac{k}{N} \right) \label{eq1}
\end{equation}
where $p_k=E_{k}/\sum_{n=1}^{N}E_{n}$ and $\omega =2 \pi \phi$,
and $\phi$ stands for the frequency in natural time, termed {\em
natural} frequency. We then compute the power spectrum
$\Pi(\omega)$ of $\Phi(\omega)$ as
\begin{equation}
\label{eq2}
\Pi(\omega)=\left| \Phi(\omega) \right|^2
\end{equation}
If we regard $p_k$  as a probability density function,
$\Phi(\omega)$ may be justified to be treated mathematically as a
characteristic function in analogy with the probability
theory\cite{FEL71}. Then, the properties of the distribution of
$p_k$ can be estimated by the expansion of this characteristic
function for $\omega\longrightarrow  0$. 

The optimality of the natural time representation of time series
has been recently shown\cite{ABE05} by means of the following
procedure: The structure of the time-frequency
representation\cite{COH94} of the signals was studied by employing
the Wigner function\cite{WIG32} to compare the natural time
representation with the ones, either in conventional time or other
possible time reparametrizations. Significant enhancement of the
signal was found in the time-frequency space if natural time is
used, in marked contrast to other time domains. Since in time
series analysis, it is desired to reduce uncertainty and extract
signal information as much as possible, the most useful time
domain should maximize the information measure, and hence minimize
the entropy. This was statistically ascertained in natural time,
by investigating a multitude of different time domains in several
electric signals.

Natural time $\chi$, from its definition, is {\em not continuous}
and takes values which are {\em rational} numbers in the range
(0,1]. (In these numbers, as the complex system evolves, the
numerators are just the natural numbers (except 0), which denote
the {\em order} of the appearance of the consecutive events).
 Hence, one
of the fundamental differences between (conventional) time and
natural time refers to the fact that the former is based on the
idea of {\em continuum}, while the latter is {\em not}. This paper
aims at raising some consequences of this difference, and in
particular those that stem from the {\em set theory} developed by
Cantor, having in mind the following remark made by
Schr\"{o}dinger (see pp. 62-63 of Ref.\cite{SCH54}):

``We are familiar with the idea of {\em continuum}, or we believe
ourselves to be. We are {\em not} familiar with the enormous
difficulty this concept presents to the mind, unless we have
studied very modern mathematics (Dirichlet, Dedekind,Cantor).''

We clarify in advance that we do not tackle here the case (since
it is inapplicable to our universe\cite{HEN05}) raised by
G\"{o}del in 1949 who discovered\cite{GOD49} unexpected solution
to the equations of general relativity corresponding to universes
in which no universal temporal ordering is possible(see also
Refs.\cite{DAV05A,YOU05} and references therein). This solution
acquires its simplest form (see p.86 in \cite{SCH85}) ``with {\em
two} of the coordinate-line-elements time-like (the other two
space-like)''. Interestingly, Schr\"{o}dinger in an early version
of Ref.\cite{SCH85}, which was published almost simultaneously
with G\"{o}del's work, had also emphasized that ``there is no
necessity for just three of the four line-elements being
space-like, one time-like \dots '' (cf. note also that 
very recently it was suggested\cite{HAW06} that when the universe was small
enough to be governed by quantum mechanics, it had four spatial
dimensions and {\em no} dimension of time.)

We now recapitulate some points of the Cantor's set theory that
are relevant to our present discussion. A {\em transfinite number
or transfinite cardinal} is the cardinality of some {\em infinite}
set, where the term {\em cardinality} of a set stands for the
number of members it contains, e.g., pp. 2-3 of \cite{SUB98}. The
set of natural numbers is labeled by $N$, while the number of
natural numbers is designated by $\aleph_0$, i.e., $\aleph_0 =
|N|$ (cf. the cardinality of a set $S$ is labelled $|S|$). In this
transfinite number, the zero subscript is justified by the fact
that, as proved by Cantor, no infinite set has a smaller
cardinality than the set of natural numbers. It can be shown that
the set of rational numbers designated by $Q$ has the same
cardinality as the set of natural numbers, or $|N| = |Q|$ (e.g.,
Theorem 2 in \cite{SUB98}). In other words, the rationals are {\em
exactly} as numerous as the naturals. Note that a set is {\em
countable} iff its cardinality is either finite or equal to
$\aleph_0$ and in particular is termed {\em denumerable} iff its
cardinality is exactly $\aleph_0$. (cf. As usually, for ``if and
only if'' we write simply ``iff''). A set is {\em uncountable} iff
its cardinality is greater than $\aleph_0$, see also below. Hence,
natural time takes values (which, as mentioned, are rational
numbers) that form in general  a {\em countable} set; this becomes
a {\em denumerable} set\cite{EPAPS} if the number of events is infinitely
large. Further, since in natural time analysis we consider the
pairs $(\chi_k,E_k)$, the values of the quantity $E_k$ {\em
should} form a set with cardinality equal to (or smaller than)
$\aleph_0$. In other words, the values of the energy also form a
{\em countable} set, which reflects of course that the energy is
{\em not continuous}.

The fact that $|N| = |Q|$ is an astounding result in view of the
following: The rational numbers are {\em dense} in the real
numbers, which means that between any two rational numbers on the
real number line we can find {\em infinitely more} rational
numbers. In other words, although the set of rational numbers
seems to contain infinities within infinities, there are just as
many natural numbers as there are rational numbers. This reflects
the following point: Let us assume that we follow a system with
some (experimental) accuracy, in which, as mentioned, after an
infinite number of events the cardinality of the set of the values
of natural time is $\aleph_0$. Let us assume that we now repeat
the measurement with more sensitive instrumentation (i.e.,
counting events above an appreciably smaller energy or duration
threshold in Figs. 1b and 1a, respectively) and hence between two
consecutive events of the former measurement a considerable number
of appreciably smaller events is monitored. The corresponding
cardinality, in contrast to our intuition, is again $\aleph_0$.
(The inverse, i.e., when the instrumentation becomes less
sensitive, may correspond to a ``coarse graining'' procedure). In
others words, when considering the occurrence of infinitely large
number of consecutive events, the natural time takes values that
form a {\em denumerable} set and this remains so even upon
increasing the accuracy (and hence lowering the uncertainty) of
our measurement.

We now turn to the aspects of Cantor's set theory related to the
real numbers, which as mentioned are associated with the
conventional time. It is shown that the number of points on a
finite line segment is the same as the number of points on an
infinite line (e.g., Theorem 13 in \cite{SUB98}). Considering the
definition: The number of real numbers is the same as the number
of points on an infinite line (or in the jargon, the {\em
numerical continuum} has the same cardinality as the {\em linear
continuum}), let ``c'' designate the cardinality of the continuum
-or equivalently the cardinality of the set of real numbers.
(Hence $c=|R|$  by definition). It is proven (e.g., Theorem 16 in
\cite{SUB98}) that the set of real numbers is uncountable, or $|R|
> \aleph_0$. (Equivalently, this theorem asserts that $c >
\aleph_0$). Hence, the values of conventional time form an {\em
uncountable} set, in contrast to that of natural time which as
mentioned is countable. In order to further inspect this
fundamental difference, we resort to the continuum hypothesis (CH)
-see below- which was formulated (but not proved) by Cantor. This,
after Euclid's parallel postulate, was the first major conjecture
to be proved {\em undecidable} by standard
mathematics\cite{EPAPS}.

We first clarify that the power set $^*S$ of a set $S$, which is
the set of all subsets of $S$, has a cardinality $|^*S|=2^{|S|}$
when $S$ is finite. According to Cantor's Theorem the cardinality
of the power set of an {\em arbitrary} set has a greater
cardinality than the original arbitrary set, i.e., $|^*S|>|S|$
(e.g., Theorem 4 in \cite{SUB98}). This theorem is trivial for
finite sets, but fundamental for infinite sets. Hence, for any
infinite cardinality, there is a larger infinite cardinality,
namely, the cardinality of its power set. From CH, which asserts
that there is {\em no} cardinal number $\alpha$ such that
$\aleph_0 < \alpha < c$, it follows that the next largest
transfinite cardinal after $\aleph_0$ (labelled $\aleph_1$) is
$c$, thus $c = \aleph_1$. Since Cantor proved (e.g., Theorem 17 in
\cite{SUB98}) that $\aleph_1=2^{\aleph_0}$, CH leads to: $c =
2^{\aleph_0}$ (thus, this is the number of points on an infinite
line). Hence, if we assume CH, the cardinality of the set of the
values of natural time -when considering infinitely large number of
events- corresponds to $\aleph_0$, while that of the conventional time
is $2^{\aleph_0}$. The values of the former, as mentioned, are
rational numbers, while almost all the values of the latter are
{\em irrational}, because, since $2^{\aleph_0} \gg \aleph_0$,
almost all reals are irrational numbers. (On the other hand,
without assuming CH we have essentially no idea which transfinite
number corresponds to $c$, and we would know the cardinality of
the naturals, integers, and rationals, but {\em not} the
cardinality of the reals, e.g., p. 15 of \cite{SUB98}). As for the
values of $E_k$, they are {\em not} necessarily
rational, because in general when taking $\aleph_0$ (at the most)
out of $2^{\aleph_0}$ values they may all be irrational. Hence,
even upon gradually improving the accuracy of our measurements,
both sets $\{ \chi_k \}$ and $\{ E_k \}$ remain
denumerable, the former consisting of rational numbers only. 

Schr\"{o}dinger, in order to point out the ``Intricacy of the
continuum'', used an example analogous to the Cantor set $C$ (see
pp. 138-143 of Ref.\cite{SCH51}). The latter is given by taking
the interval [0,1], removing the open middle third, removing the
middle third of each of the two remaining pieces, and continuing
this procedure ad infinitum. The cardinality of this set $C$ is no
less than that of [0,1]. Since $C$ is a subset of [0,1], its
cardinality is also no greater, so it must in fact be equal. In
other words, there are as many points in the Cantor set as there
are in [0,1], and the Cantor set is uncountable. The same holds
for the {\em random} Cantor set, which has been suggested as being
involved in the geometrical description of the fluctuations of the
vacuum\cite{EL04} (cf. modern physicists hypothesize that what
appears to our senses as empty space is in reality a richly
dynamical medium \cite{WIL05C,WIL06}, which has energy, e.g.,
\cite{VIL06,CAR06}.). Hence the cardinality of either a Cantor set
or a random Cantor set differs drastically from that of the set of
the values of natural time.

We finally comment on the common view that (conventional) time is
continuous, keeping in the frame that, as pointed out by
Schr\"{o}dinger (p. 145 of Ref.\cite{SCH58}) ``our sense
perceptions constitute our sole knowledge about things''. In
short, it seems that the continuity of time does not stem from
{\em any} fundamental principle, but probably originates from the
following general demand on continuity discussed by
Schr\"{o}dinger (see p. 130 of \cite{SCH51}):

``From our experiences on a large scale........ physicists had
distilled the one clear-cut demand that a truly clear and complete
description of any physical happening has to fulfill: it ought to
inform you precisely of what happens at any point in space at any
moment of time ....... We may call this demand the {\em postulate
of continuity of the description}''. Schr\"{o}dinger, however,  subsequently commented on this demand as follows
(see p.131 of \cite{SCH51}): ``It is this postulate of continuity
that appears to be unfulfillable!......'' and furthermore added:
``We {\em must not admit the possibility of continuous
observation}''. If we attempt a generalization of these intuitive
remarks, we may say that the concept of natural time is not
inconsistent with Schr\"{o}dinger's point of view.

In summary, conventional time is currently assumed continuous, but
this does not necessarily result from any fundamental principle.
Its values form an uncountable set, almost all of which are {\em
irrational} numbers. On the other hand, natural time is not
continuous, and its values form a countable set consisting of
rational numbers only; further, the values of the energy, which
are not necessarily rational, also form a countable set. Upon
considering an infinitely large number of events, the cardinality
of the set of the values of natural time is $\aleph_0$ (cf. it
persists even upon increasing the accuracy of the measurement),
thus being drastically smaller than that of conventional time,
which equals to $2^{\aleph_0}$ if we accept the validity of  the continuum
hypothesis.

\bibliographystyle{apsrev}

\end{document}